\documentclass[12pt, a4paper]{article}

\usepackage{graphicx}

\usepackage{stmaryrd}
\usepackage{overcite}
\usepackage{verbatim}
\usepackage{achemso}
\usepackage{verbatim}

\usepackage{multirow}
\usepackage{amsmath}
\usepackage{amsfonts}
\usepackage{amsgen}
\usepackage{longtable}
\usepackage{lscape}
\usepackage{threeparttable}
\usepackage{supertabular}
\usepackage{multicol}
\usepackage{float}
\usepackage{multirow}
\usepackage{setspace}
\usepackage{rotating}
\usepackage{amssymb}
\usepackage{layout}
\usepackage{wrapfig}
\usepackage{subfig}
\usepackage{enumerate}
\usepackage{subfig}
\usepackage{color}
\usepackage{soul}

\begin{document}

\begin{center}

{\bf\large{Interdigitated ring electrodes: Theory and experiment}}

\hspace{2cm}

{\bf\large Edward O. Barnes$^{a}$, Ana Fern\'{a}ndez la Villa$^{b}$, Diego F. Pozo Ayuso$^{b}$, Mario Casta\~{n}o Alvarez$^{b}$, Grace E. M. Lewis$^{c}$, Sara E. C. Dale$^{c}$, Frank Marken$^{c}$ and Richard G. Compton$^{a}$*}

*Corresponding author\\
Email:~richard.compton@chem.ox.ac.uk\\
Fax:~+44~(0)~1865~275410; Tel:~+44~(0)~1865~275413.\\

$^{a}$Department~of~Chemistry, Physical~and~Theoretical~Chemistry~Laboratory, Oxford~University,
South~Parks~Road, Oxford, OX1~3QZ, United~Kingdom.\\
$^b$MicruX~Fluidic, S.L., Edificio~Severo~Ochoa, Juli\'{a}n~Claver\'{i}a, s/n, 33006~Oviedo~(Asturias), Spain.\\
$^c$Department~of~Chemistry, University~of~Bath, Bath, BA2~7AY, United~Kingdom.
\vspace{1cm}

To be submitted as an article to:\\
\emph{The Journal of Electroanalytical Chemistry}

\end{center}

\clearpage

\section*{Abstract}

The oxidation of potassium ferrocyanide, K$_4$Fe(CN)$_6$, in aqueous solution under fully supported conditions is carried out at interdigitated band and ring electrode arrays, and compared to theoretical models developed to simulate the processes. Simulated data is found to fit well with experimental results using literature values of diffusion coefficients for Fe(CN)$_6^{4-}$ and Fe(CN)$_6^{3-}$. The theoretical models are used to compare responses from interdigitated band and ring arrays, and the size of ring array required to approximate the response to a linear band array is investigated. An equation is developed for the radius of ring required for a pair of electrodes in a ring array to give a result with 5\% of a pair of electrodes in a band array. This equation is found to be independent of the scan rate used over six orders of magnitude.

\section*{Keywords}

electrochemical simulation, electrode arrays, generator-collector systems, interdigitated band electrodes, interdigitated ring electrodes

\clearpage

\section{Introduction}

Microelectrode arrays are increasingly employed in electrochemistry\cite{Ordeig2007, Herzog2013}, since they can provide the advantages of microelectrodes (fast mass transport, the use of two rather than three electrode control, observation of steady state currents) whilst providing the large currents usually associated with macroelectrodes. These arrays also have considerably less capacitative current than macroelectrodes, although generally more than isolated microelectrodes, representing an intermediate case\cite{Fletcher1999}. However, while single microelectrodes typically have a much smaller ohmic drop than macroelectrodes (due to the smaller currents drawn), a microelectrode array has significantly larger ohmic drop than a macroelectrode of equivalent overall area\cite{Ordeig2007}. The main types of array electrodes are arrays of microdiscs (where the array may be regular or random) and arrays of microbands.

Microband arrays, as well as a single pair of bands, have been used as generator-collector systems\cite{Tomcik1997, Sanderson1985, Jencusova2006, Aoki1988, Aoki1989, Strutwolf2005, Barnes2012, Rajantie2001a, Rajantie2001}, with great success. In a collector generator system, some target species in solution is oxidised (or reduced) at the generator electrode, with the product of this reaction then detected by re-reduction (or oxidation) at the collector electrode. If single step chronoamperometry, for example, were to be carried out, then some species might would be oxidised at the generator electrode to produce species B, which would then be reduced at the collector electrode, both under diffusion control:
\begin{eqnarray}
\mathrm{A} \pm \mathrm{e}^- & \rightarrow & \mathrm{B} \quad \text{Generator}\\
\mathrm{B} \mp \mathrm{e}^- & \rightarrow & \mathrm{A} \quad \text{Collector}
\end{eqnarray}
An important feature in an experiment of this type is the collection efficiency, $N$, the ratio of the collector current to the generator current:
\begin{equation}
N = -\frac{I_\mathrm{col}}{I_\mathrm{gen}}
\end{equation}
This simple experiment can be extended to performing cyclic voltammetry at the generator electrode, while keeping the collector electrode fixed at the potential to reduce species B. This approach has been used to study, for example, competing modes of transport at a phase boundary (transport along the interface vs transport in bulk)\cite{French2009b}. Other uses of collector generator systems include mechanistic invesigations\cite{Nekrasov1967a, Damjanovic1967, Unwin1991, Bruckenstein1977}, probing intermediate kinetics\cite{Albery1966d, Albery1966e}, electrochemical sensing with very low detection limits\cite{Dale2011}, targeted detection of one species in the presence of another\cite{Zhu2011}, and the simultaneous measurement of two species\cite{Barnes2013a}.

\begin{figure}[htb]
\centering
\subfloat[IDBA]{\includegraphics[width=0.4\textwidth]{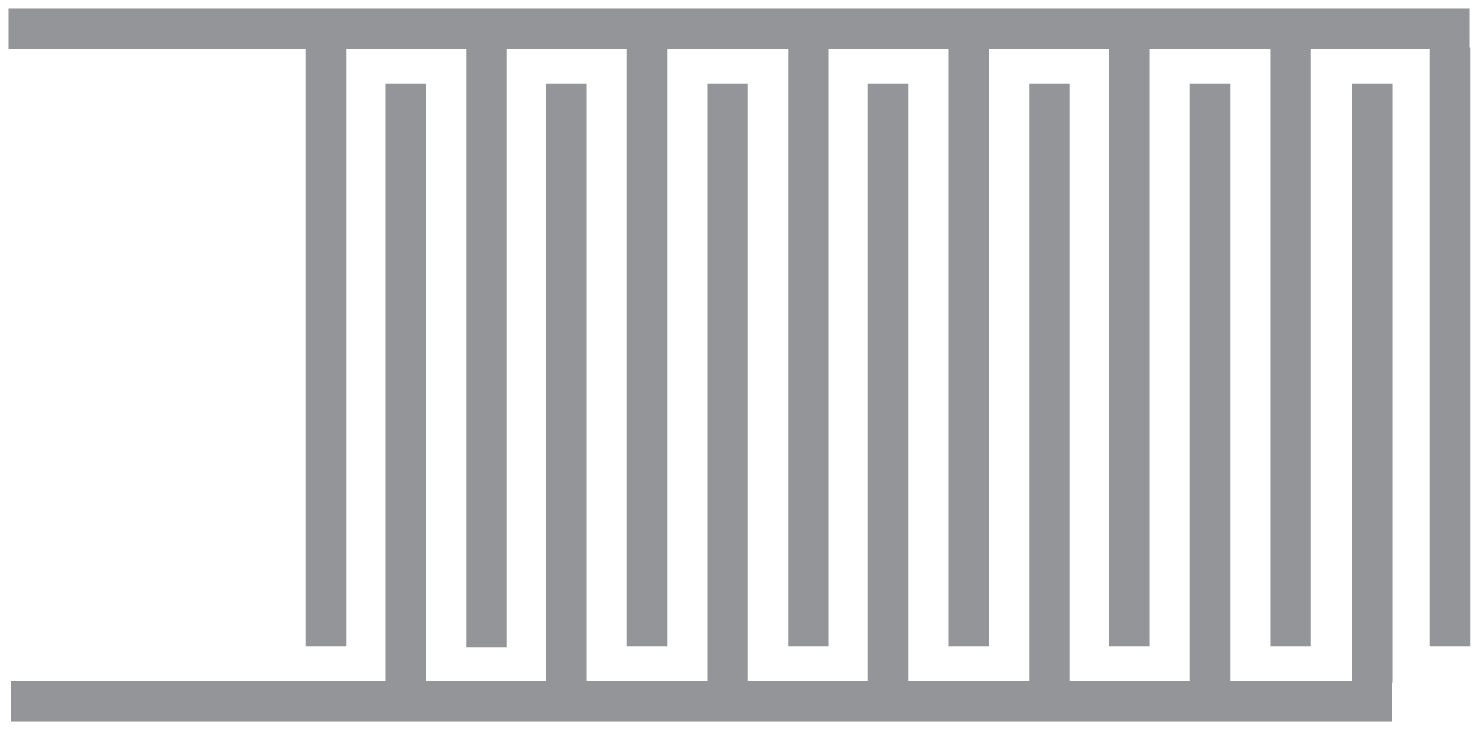}\label{IDA SCHEMATIC}}\hspace{20pt}
\subfloat[IDRA]{\includegraphics[width=0.4\textwidth]{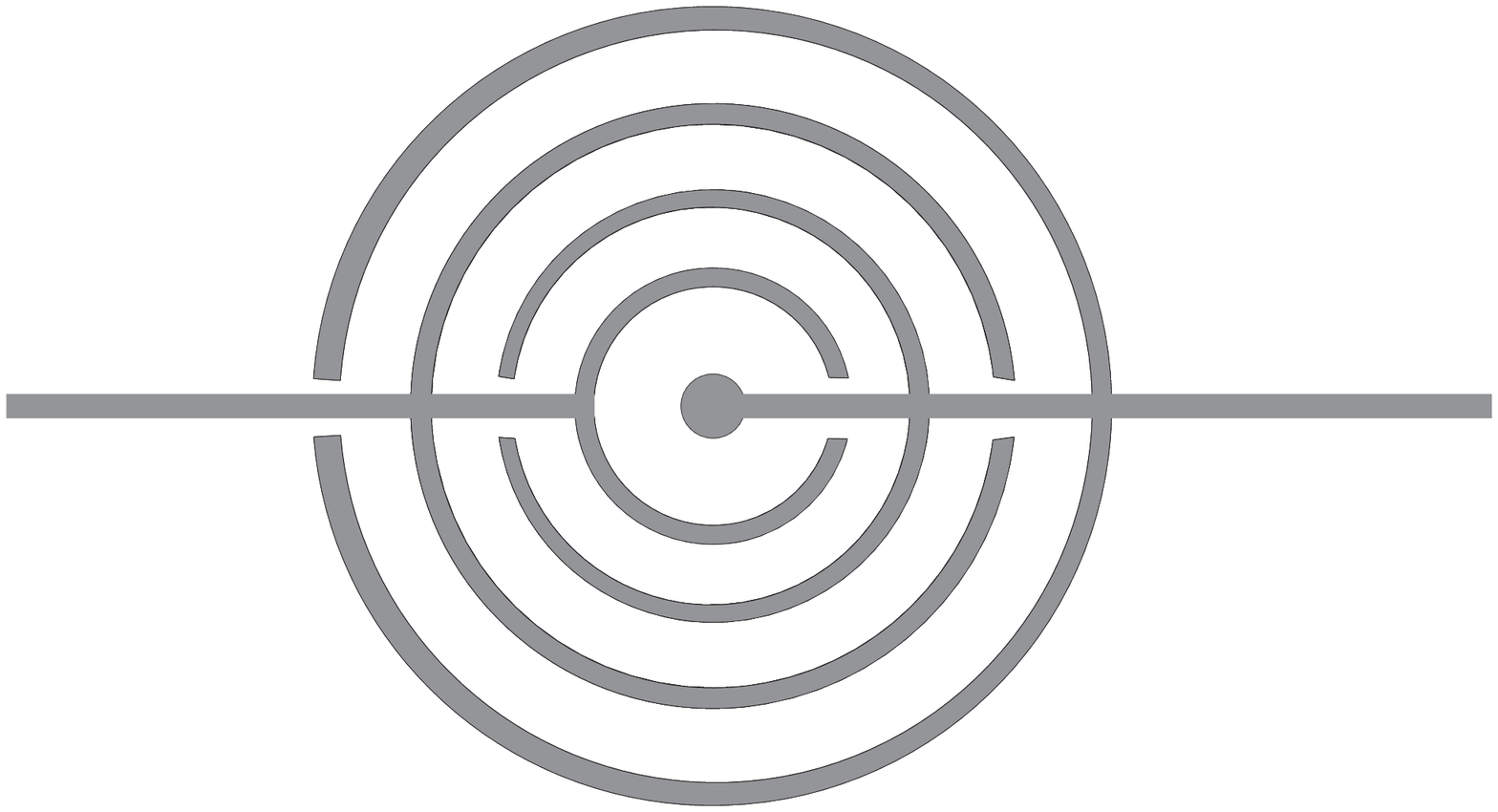}\label{IDRA SCHEMATIC}}
\caption{Schematic diagram of the two interdigitated arrays investigated in this study.}
\end{figure}

Microband arrays lend themselves particularly well to generator-collector systems. They consist of a series of parallel bands, which alternate between being generator and collector. This is shown in Fig. \ref{IDA SCHEMATIC}, and the apparatus is known as an interdigitated array (IDBA). This ensures a very high collection efficiency\cite{Niwa1990}, since any species produced at the generator electrode is likely to diffuse to the collector electrode with a high probability of detection if the separation is small. This spatial arrangement also allows for efficient redox cycling\cite{Niwa1990}, where species cycle between generator and collector electrodes, enhancing the measured current.

IDBAs have been used to simultaneously measure both the concentration and diffusion coefficient of a species. Aoki \emph{et. al.}\cite{Aoki1989} developed a theoretical equation for the steady state current produced at a parallel microband array electrode during a chronoamperometric experiment as outlined above:
\begin{equation}
I_{SS} = nFDc^*l\frac{K(1-p)}{K(p)} \label{EQUATION1}
\end{equation}
where $I_{ss}$ is the steady state current (A), $n$ is the number of electrons transfered, $F$ is the Faraday constant, $D$ is the diffusion coefficient of the species under investigation (m$^2$~s$^{-1}$), $c^*$ is this species' bulk concentration (mol~m$^{-3}$), $l$ is the length of the individual band electrodes (m), and $p$ is some function of the electrode geometry. The function $K$ is a total elliptic integral, and $K(1-p)/K(p)$ can be approximated as:
\begin{equation}
\frac{K(1-p)}{K(p)} \approx 0.318\mathrm{ln}\left[2.55\left(1 + \frac{w_e}{w_g}\right)\right] - 0.095\left(1 + \frac{w_e}{w_g}\right)^{-2}
\end{equation}
where $w_g$ and $w_e$ are the inter-electrode distance and the electrode width respectively. Also developed was an empirical equation to described the time taken for the current at the collector electrode to reach half of its steady state value, $t_{0.5}$:
\begin{equation}
t_{0.5} \approx 0.90\frac{\left(w_g + \frac{w_e}{6}\right)^2}{D} \label{EQUATION2}
\end{equation}
So by measuring the steady state current and the time taken for to reach half this value, equations \ref{EQUATION1} and \ref{EQUATION2} can be solved simultaneously to determine $c^*$ and $D$.

A convenient and important variation on IDBAs are interdigitated ring electrode arrays, IDRAs\cite{Niwa1996}. a schematic of an IDRA is shown in Fig. \ref{IDRA SCHEMATIC}. They consist of a central disc electrode (throughout this paper considered to be a generator electrode) surrounded by expanding concentric ring electrodes which alternate between collector and generator in character. The obvious difference between this and an IDBA is the cylindrical nature of the diffusion. The radial component will, at least for rings of small radii, enhance the mass transport between the electrodes, leading to an increased rate of redox cycling and a greater current enhancement in comparison with linear IDBAs. For very large rings, however, the diffusion will be become less cylindrical and these effects will be lost, with the behavior tending to that of parallel bands. This study investigates the extent to which this is the case, and explores how well experiments carried out at an IDRA can be simulated a simplified IDBA model.

\section{Theory}

Cyclic voltammetry is simulated at both linear interdigitated arrays (IDBAs) and interdigitated ring arrays (IDRAs), shown in Fig \ref{IDA SCHEMATIC} and \ref{IDRA SCHEMATIC} respectively. In each case, a single electroactive species, A, is considered to be initially present in solution, along with a large amount of an inert electrolyte to suppress electric fields and render the mass transport diffusion only. The generator electrodes are subject to a linearly sweeping potential to set up a potential dependent equilibrium between species A and its oxidised form, species B, while the collector electrodes are set at a fixed potential to reduce species B at a mass transport controlled rate:
\begin{eqnarray}
\mathrm{A} \pm \mathrm{e}^- & \rightleftharpoons & \mathrm{B} \quad \text{Generator} \label{GENERATOR}\\
\mathrm{B} \mp \mathrm{e}^- & \rightarrow & \mathrm{A} \quad \text{Collector} \label{COLLECTOR}
\end{eqnarray}

\subsection{Interdigitated arrays}

\begin{figure}[h]
\begin{center}
\includegraphics[width = 0.9\textwidth]{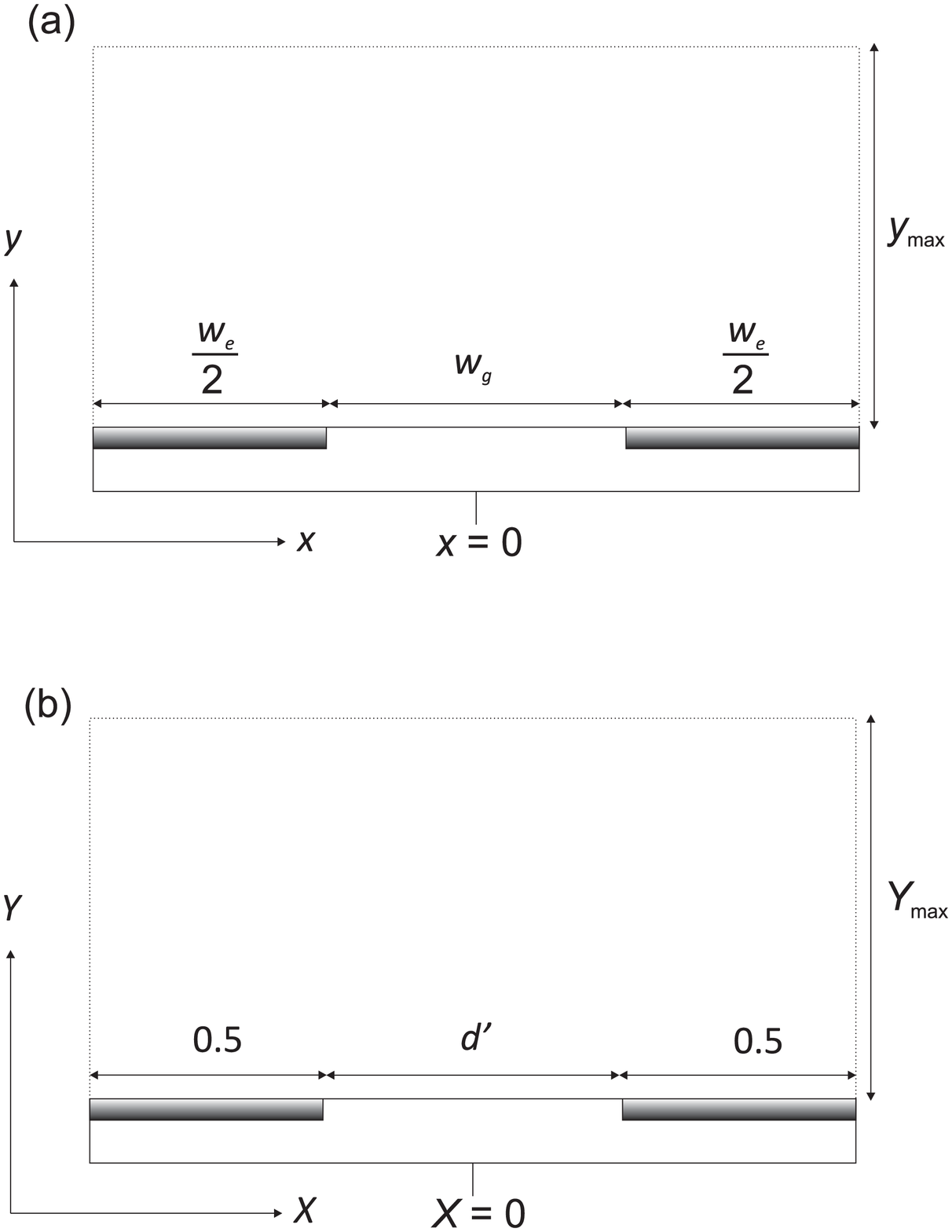}
\caption{(a) Dimensional and (b) normalised simulation space used in the IDA model.} \label{IDA SIM SPACE}
\end{center}
\end{figure}

To simulate a linear interdigitated array, we make the assumption that the electrodes are much longer than they are wide, reducing the problem from three dimensions to two. The mass transport equation is therefore:
\begin{equation}
\frac{\partial{c_\mathrm{i}}}{\partial{t}} = D_\mathrm{i}\left(\frac{\partial^2{c_\mathrm{i}}}{\partial{x}^2} + \frac{\partial^2{c_\mathrm{i}}}{\partial{y}^2}\right)
\end{equation}
where $c_\mathrm{i}$ is the concentration of species i (mol~m$^{-3}$) and $D_\mathrm{i}$ is the diffusion coefficient of species i (m$^2$~s$^{-1}$). All symbols are defined in Table \ref{DIMENSIONAL}. The simulation space which this is solved over is shown schematically in Fig. \ref{IDA SIM SPACE}a. The mass transport equation is solved subject to boundary conditions. At the generator electrode surface, Butler-Volmer kinetics are applied:
\begin{equation}
D_\mathrm{A}\left(\frac{\partial{c_\mathrm{A}}}{\partial{y}}\right) = k^0\left[c_\mathrm{A}^0\text{exp}\left(\frac{-\alpha\left(E - E_f^\minuso\right)}{RT}\right) - c_\mathrm{B}^0\text{exp}\left(\frac{\left(1-\alpha\right)\left(E - E_f^\minuso\right)}{RT}\right)\right]
\end{equation}
and equal fluxes of species are maintained by virtue of conservation of mass:
\begin{equation}
D_\mathrm{B}\left(\frac{\partial{c_\mathrm{B}}}{\partial{y}}\right) = -D_\mathrm{A}\left(\frac{\partial{c_\mathrm{A}}}{\partial{y}}\right)
\end{equation}
At the collector electrode surface, the potential is such that species B is reduced to species A and a mass transport controlled rate, and equal fluxes are again maintained:
\begin{eqnarray}
c_\mathrm{B} & = & 0\\
D_\mathrm{A}\left(\frac{\partial{c_\mathrm{A}}}{\partial{y}}\right) & = & -D_\mathrm{B}\left(\frac{\partial{c_\mathrm{B}}}{\partial{y}}\right)
\end{eqnarray}
The edges of the simulation space are set at the centres of the electrodes ($x_\mathrm{min} = -\frac{1}{2}w_g - w_e$ and $x_\mathrm{max} = \frac{1}{2}w_g + w_e$) a distance along the y axis from the electrodes known to be well outside the diffusion layer\cite{Gavaghan1998b, Svir2001} ($y_\mathrm{max} = 6\sqrt{D_\mathrm{max}t_\mathrm{max}}$) where $D_\mathrm{max}$ and $t_\mathrm{max}$ are the largest diffusion coefficient in the system, and the total time of the experiment. At the simulation edges, a zero flux condition is imposed on all species:
\begin{eqnarray}
\left(\frac{\partial{c_\mathrm{i}}}{\partial{x}}\right)_{x_\mathrm{min}\text{, }x_\mathrm{max}\text{, all }y} = 0\\
\left(\frac{\partial{c_\mathrm{i}}}{\partial{y}}\right)_{\text{all }x\text{, }y_\mathrm{max}} = 0\\
\end{eqnarray}
The value of the potential applied to the generator electrode at a given time from the start of the experiment must also be defined. If the scan starts at some potential value $E_\mathrm{s}$ (V), and sweeps at a scan rate of $\nu$ (V~s$^{-1}$) up to a vertex potential $E_\mathrm{v}$ (V) before reversing and scanning back to $E_\mathrm{s}$, the the applied potential $E$ at any given time $t$ is:
\begin{equation}
E = E_\mathrm{v} \pm |E_\mathrm{v} - E_\mathrm{s} \pm \nu t|
\end{equation}
where $+$ is used for a reduction and $-$ for an oxidation.

\subsubsection{Normalised Model}

The above model is next simplified by introducing normalised (or ``dimensionless'') variables to make it general. The bulk concentration of species A, for example, is merely a scaling factor and is set to 1, with all other concentrations calculated relative to this. Similarly, all diffusion coefficients are set relative to $D_\mathrm{A}$, and all distances relative to the width of the electrode. A full list of normalised parameters and their definition is given in Table \ref{DIMENSIONLESS}. The normalised simulation space is shown schematically in Fig. \ref{IDA SIM SPACE}b. Upon the substitution of dimensional parameters for normalised ones, the mass transport equation becomes:
\begin{equation}
\frac{\partial{C_\mathrm{i}}}{\partial{\tau}} = D^{'}_\mathrm{i}\left(\frac{\partial^2{C_\mathrm{i}}}{\partial{X}^2} + \frac{\partial^2{C_\mathrm{i}}}{\partial{Y}^2}\right)
\end{equation}
The normalised boundary conditions are summarised in Table \ref{IDA BOUNDARY CONDITIONS}. The normalised potential applied to the generator ($\theta$) electrode is still a function of normalised scan rate, $\sigma$:
\begin{equation}
\theta = \theta_\mathrm{v} \pm |\theta_\mathrm{v} - \theta_\mathrm{s} \pm \sigma\tau|
\end{equation}
again with $+$ used for a reduction and $-$ for an oxidation.

\subsubsection{Calculating the current at an IDBA}

Upon the implementation of the above model, the current at both the generator and the collector electrodes must be calculated. The (dimensionless) flux density at an individual generator electrode is given by integrating the flux across its width:
\begin{equation}
j_\mathrm{gen} = 2\int_{-0.5 - \frac{1}{2}d}^{- \frac{1}{2}d}{D^{'}_\mathrm{A}\left(\frac{\partial{C_\mathrm{A}}}{\partial{Y}}\right)_{Y=0}} \, \mathrm{d}X
\end{equation}
and at the collector:
\begin{equation}
j_\mathrm{col} = 2\int_{\frac{1}{2}d}^{0.5 + \frac{1}{2}d}{D^{'}_\mathrm{A}\left(\frac{\partial{C_\mathrm{A}}}{\partial{Y}}\right)_{Y=0}} \, \mathrm{d}X
\end{equation}
Note the factors of 2, these are due to using the symmetry of an interdigitated array to simulate only half of an individual generator and collector electrode. The dimensional current is then given by:
\begin{equation}
I = nFlmDc_\mathrm{A}^{*}j
\end{equation}
where m is the number of generator-collector pairs.

\subsection{Interdigitated ring arrays}

\begin{figure}[h]
\begin{center}
\includegraphics[width = 0.9\textwidth]{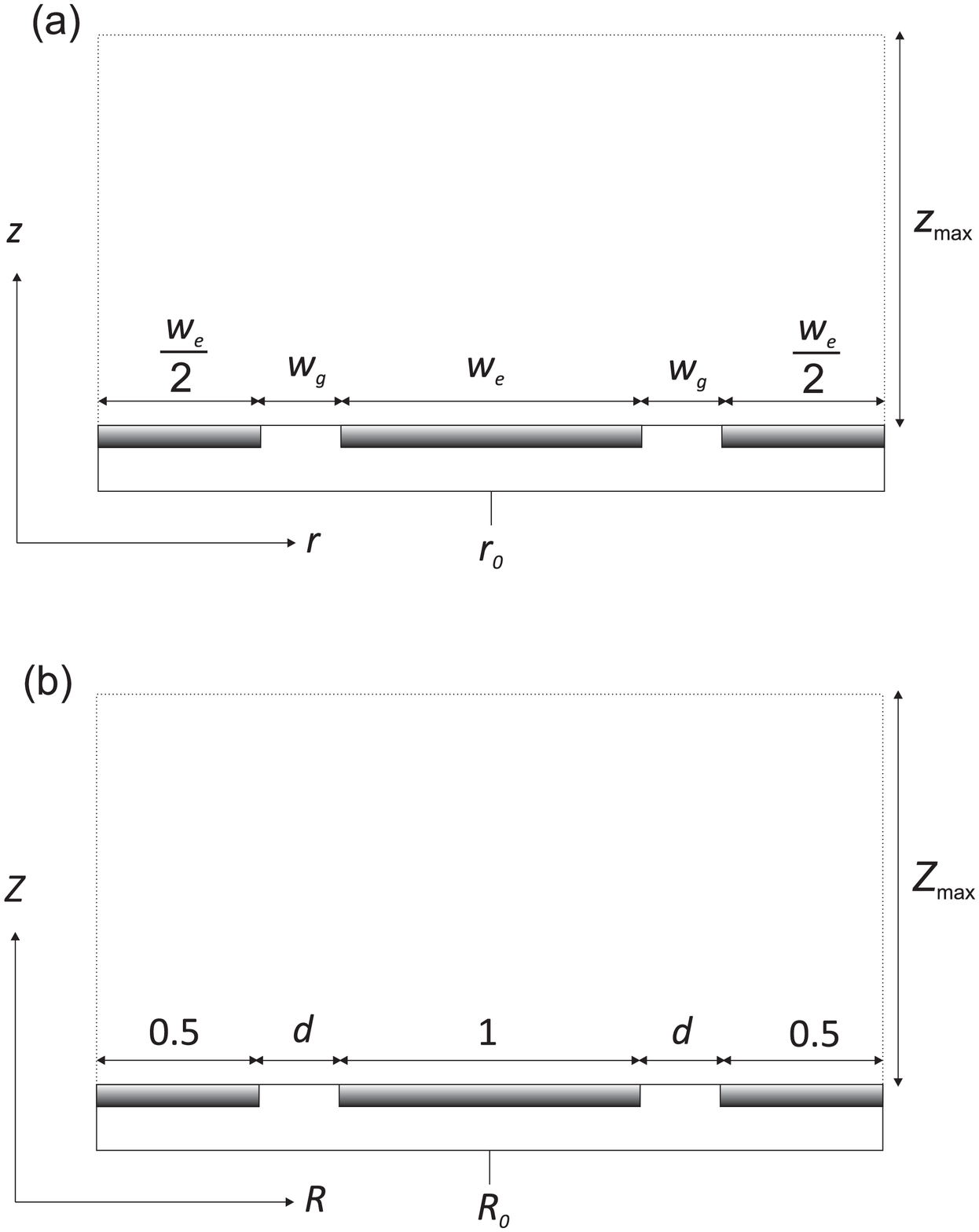}
\caption{(a) Dimensional and (b) normalised simulation space used in the IDRA model.} \label{IDRA SIM SPACE}
\end{center}
\end{figure}

To simulated an IDRA, the model developed above must be modified to account for the added radial nature of the diffusion. The new (dimensionless) mass transport equation is:
\begin{equation}
\frac{\partial{C_\mathrm{i}}}{\partial{\tau}} = D^{'}_\mathrm{i}\left(\frac{\partial^2{C_\mathrm{i}}}{\partial{R}^2} + \frac{1}{R}\frac{\partial{C_\mathrm{i}}}{\partial{R}}  + \frac{\partial^2{C_\mathrm{i}}}{\partial{Z}^2}\right)
\end{equation}
where $R$ is the dimensionless radial coordinate and $Z$ is the dimensionless coordinate perpendicular to the surface of the electrodes, as defined in Table \ref{DIMENSIONLESS}. Fig. \ref{IDRA SIM SPACE}a and \ref{IDRA SIM SPACE}b show schematically the dimensional and normalised simulation space this equation is to be solved over. The key difference between this simulation space and that used for the IDBA model (Fig. \ref{IDA SIM SPACE}) is that the whole width of the generator electrode is now included in the centre, with half of the two collector electrodes on either side. The central of the simulation space is now no longer zero, but set at a value $R_0$, determined by the geometry of the IDRA being simulated. Some of the symmetry that was present in the IDBA model is lost due to the radial diffusion, which will be different on different sides of the generator electrode (due to the different values of $R$). The edge of the simulation space is in the centre of the collector electrodes, but unlike above where this was a result of symmetry, it is now an approximation.

The boundary conditions are analogous to those used in the IDBA model, and are given in Table \ref{IDRA BOUNDARY CONDITIONS}. The potential applied to the generator electrode, $\theta$, varies with time and scan rate in exactly the same way as above.

\subsubsection{Calculating the current at an IDRA}

When calculating the current at an IDRA, it must be noted that since we are working in cyclindical coordinates, a factor of $R$ must be included in the integrals to calculate the overall flux. So the flux at the generator electrode is given by:
\begin{equation}
j_\mathrm{gen} = \int_{R_0 - 0.5}^{R_0 + 0.5}{D^{'}_\mathrm{A}\left(\frac{\partial{C_\mathrm{A}}}{\partial{Z}}\right)_{Z=0}\frac{R}{R_0}} \, \mathrm{d}R
\end{equation}
and at the collector:
\begin{equation}
j_\mathrm{gen} = \int_{R_0 - d - 1}^{R_0 - d -0.5}{D^{'}_\mathrm{A}\left(\frac{\partial{C_\mathrm{A}}}{\partial{Z}}\right)_{Z=0}\frac{R}{R_0}} \, \mathrm{d}R + \int_{R_0 + d + 0.5}^{R_0 + d + 1}{D^{'}_\mathrm{A}\left(\frac{\partial{C_\mathrm{A}}}{\partial{Z}}\right)_{Z=0}\frac{R}{R_0}} \, \mathrm{d}R
\end{equation}
(the factor $\frac{1}{R_0}$ is included to normalise the response with respect to $R_0$ and make for easier comparison of IBDA and IDRA responses.) The dimensional current can then be obtained from the expression:
\begin{eqnarray}
I & = & 2\pi nFw_eR_0Dc_\mathrm{A}^{*}j\\
& = & 2\pi nFr_0Dc_\mathrm{A}^{*}j
\end{eqnarray}
This will give the current measured at one pair of generator-collector electrodes, at radial coordinate $r_0$. To simulate the current measured at the entire array, many simulations must be run, each with a different value of $r_0$, and the results summed together.

\subsection{Numerical methods}

The method of Crank and Nicolson\cite{Crank1947} was used to discretise the mass transport equations and boundary conditions to allow them to be solved numerically. The alternating direction implicit method\cite{Press2007} was used, in conjunction with the Thomas algorithm\cite{Compton1988b} to efficiently solve the diagonal matrices produced.

The spatial grid the equations were solved over has been successfully employed previously to simulate an individual pair of parallel generator-collector microbands\cite{Barnes2013c}, as well as at arrays of discs\cite{Ordeig2006} and hemispheres\cite{Ward2011} In the $Y$ direction (for an IDBA, $Z$ for and IDRA), the first grid point is at the electrode/insulating surface, and has in inital step size of $\Delta$. The grid then expands away from this surface in the following manner:
\begin{equation}
Y_\mathrm{j} = \gamma Y_{\mathrm{j} - 1}
\end{equation}
The $X$ (for an IDBA, $R$ for an IDRA)grid is analogous to this, with the grid expanding away from the edges of the electrodes until it meets a grid coming in the other direction (int the centre of gaps or electrodes) or until it meets the edge of the simulation space.

The temporal grid is a regular array of points. For each unit of dimensionless potential, $\theta$, the grid is defined to have $N_\theta$ points.

Convergence studies found the following grid parameters sufficient to produce results within $1\%$ of a fully converged value: $\Delta = 8 \times 10^{-5}$, $\gamma = 1.125$, $N_\theta = 1 \times 10^{4}$. Simulations were run on an Intel (R) Xeon (R) 2.26 GHz PC with 2.25 GB RAM, with a typical running time of 20 mins per simulation.

\section{Theoretical Results}

\begin{figure}[h]
\begin{center}
\includegraphics[width = 0.9\textwidth]{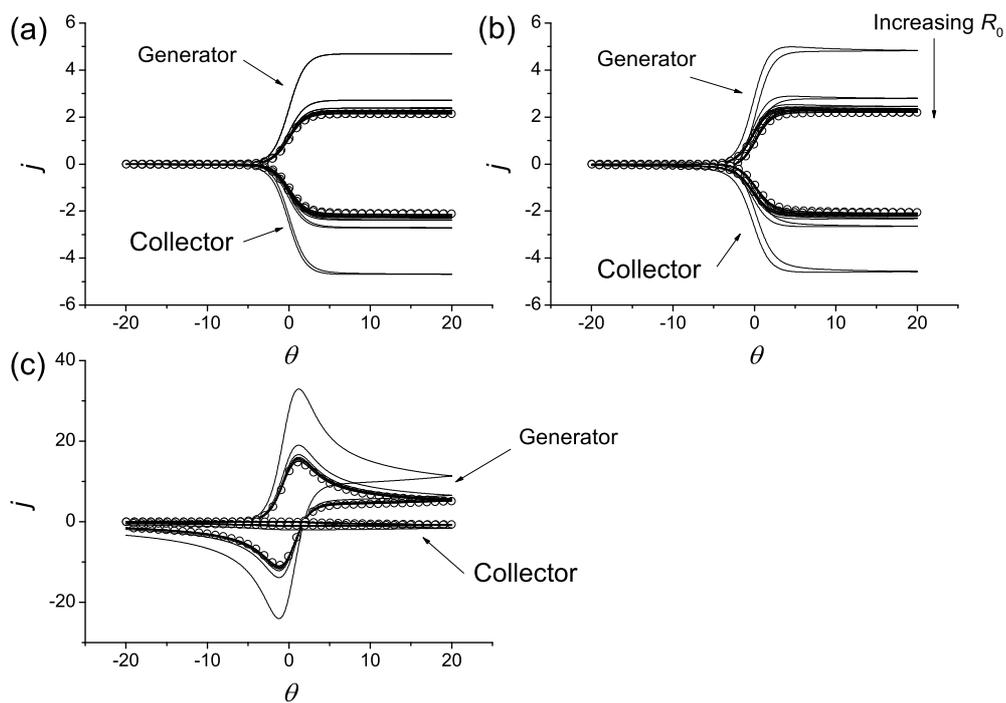}
\caption{Simulated cyclic voltammograms for a one electron oxidation at a pair of electrodes in an IDBA (circles) and a pair in an IDRA at various values of $R_0$ (= 2, 5, 10, 15, 20, 25, 50 and 100, lines). In all cases, $D^{'}_\mathrm{B} = 1$, $\alpha = 0.5$, $K^0 = 1000$. (a): $\sigma = 0.001$, (b): $\sigma = 1$ and (c): $\sigma = 1000$} \label{R AND SIGMA}
\end{center}
\end{figure}

In order to compare and contrast cyclic voltammetry at IDBAs and IDRAs, simulations were carried out at both geometries. A simple, fully electrochemically reversible one electron oxidation was simulated at a generator-collector pair within an IDBA and compared to the same reaction simulated at a generator-collector pair in an IDRA with various values of $R_0$. The results for $\sigma$ values of 0.001 (microelectrode/slow scan rate), 1 (intermediate) and 1000 (macroelectrode/fast scan rate) are shown in Fig. \ref{R AND SIGMA}. In all cases, $d^{'}$ is fixed at 0.1, $D^{'}_\mathrm{B} = 1$, $\alpha = 0.5$ and $K^0$ = 1000. It is seen that for all scan rates, a smaller value of $R_0$ enhances the dimensionless current, due to the greater radial diffusion and more efficient mass transport. As $R_0$ tends to infinity, the IDRA responses converge onto the response obtained at an IDBA, since the electrodes become effectively linear. As expected, the collector electrode response becomes very small at large values of sigma, since in this limit the diffusion is largely planar, so species B does not diffuse outwards towards the collector electrode.

\begin{figure}[h]
\begin{center}
\includegraphics[width = 0.9\textwidth]{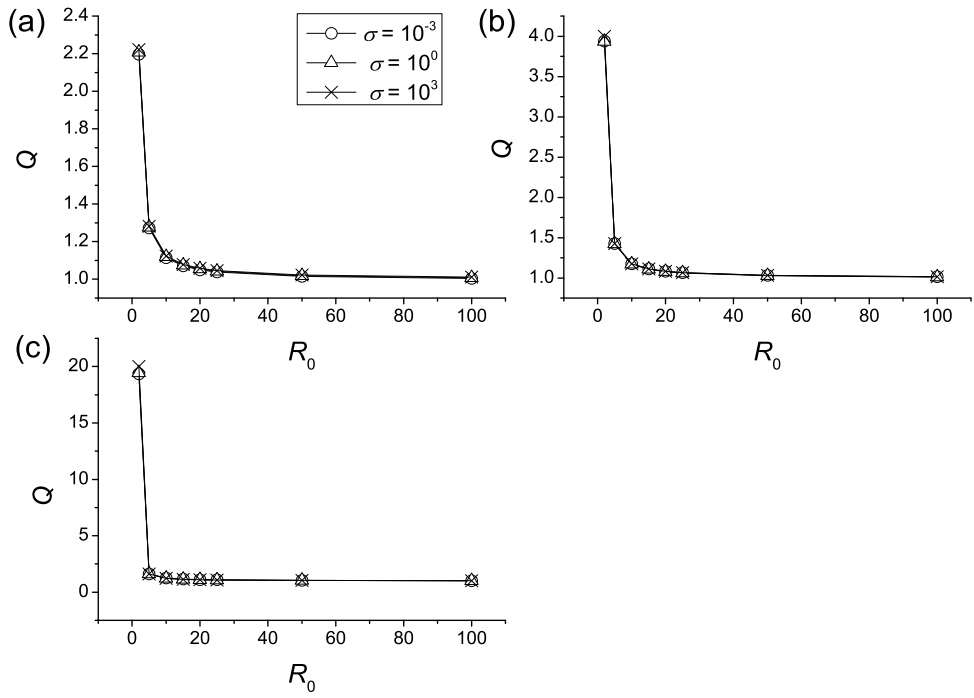}
\caption{Values of $Q$ for various $R_0$ for values of $d^{'}$ of (a): 0.1, (b): 0.5 and (c): 0.9. In each case, circles correspond to $\sigma = 0.001$, triangles to $\sigma = 1$ and crosses to $\sigma = 1000$. All other parameters as in Fig. \ref{R AND SIGMA}.} \label{SCAN RATE EFFECT}
\end{center}
\end{figure}

It will be useful to know how much the current at a pair of electrodes in an IDRA deviates from that measured at an IDBA as a function of $R_0$. This deviation, or current enhancement, is defined here as $Q$, the ratio of the peak (or steady state) generator current simulated in a IDRA simulation to that simulated at an IDBA electrode with equal $d^{'}$. From Fig. \ref{R AND SIGMA} it is seen that as $R_0$ decreases, the current enhancement ratio $Q$ increases. Fig \ref{SCAN RATE EFFECT} shows the simulated values of $Q$ for three values of $d^{'}$ (0.1, 0.5 and 0.9), each for the values of sigma (0.001, 1 and 1000). All other values remain the same as in Fig \ref{R AND SIGMA}. It is seen that at small $R_0$, $Q$ rapidly drops off as $R_0$ increases, and approaches 1 as $R_0 \to \infty$. It is also seen that, for a given value of $d^{'}$, the variation of $Q$ with $R_0$ is essentially independent of $\sigma$ over six orders of magnitude (differences between $Q$ values for any given $R_0$ is less than 1\%).

\begin{figure}[h]
\begin{center}
\includegraphics[width = 0.9\textwidth]{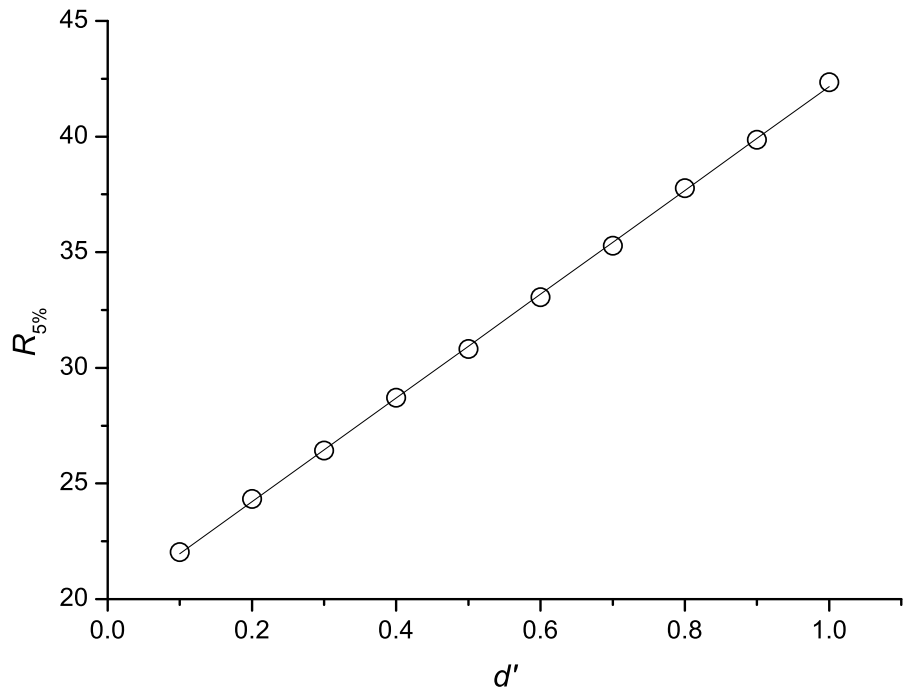}
\caption{Values of $R_{5\%}$ (the value of $R_0$ necessary for an IDRA to produce a result within 5\% of that for an IDBA) for various values of $d^{'}$.} \label{R1.05}
\end{center}
\end{figure}

Using these curves, it is then possible to calculate a value of $R_0$ at which $Q$ is a specific value, for example 1.05. Any value of $R_0$ larger than this will then produce results at an IDRA within 5\% of the reuslt for an equivalent IDBA. This value of $R_0$ is labled $R_{5\%}$. The value of $R_{5\%}$ for a given $d^{'}$ is caluclated by performing six IDRA simulations with various $R_0$ which give values of $Q$ between 1.04 and 1.06, and fitting a polynomial equation to the $Q$ vs $R_0$ curve using the Microsoft Excel graph fitting tool (a cubic equation was found to be sufficient, higher orders produced no more accurate results). This equation can then be set equal to 1.05 and solved to give a value of $R_{5\%}$. This was done for values of $d^{'}$ between 0.1 and 1.0. For each $d^{'}$, a range of scan rates between 0.001 and 1000 were used, and an avergage $R_{5\%}$ taken. The results are shown in Fig. \ref{R1.05}. It is seen that, for this range of $d^{'}$ values, $R_{5\%}$ varies linearly with $d^{'}$. A line of best fit was found using Microsoft Excel to be $R_{5\%} = 22.4d^{'} + 19.7$.

In an IDRA, the overall measured current will be dominated by the outermost electrodes, since these are largest in area. Any pair of electrodes with an $R$ coordinate at the centre of the generator greater than $R_{5\%}$ will produce currents within 5\% of those produced at an IDBA. Hence, an IDRA whose outermost electrode pair satisfies this condition is likely to be able to be simulated with reasonable accuracy as an IDBA (weighting each electrode pair's contribution according to its size).

\section{Experimental Methods}

Potassium ferrocyanide (K$_4$Fe(CN)$_6$), potassium chloride (KCl) and nitrogen gas (N$_2$) were purchased from Sigma Aldrich (USA) and used as received, without purification. Solutions were prepared in deionised water of resistivity of no less than 18.2 M$\Omega$ cm (Millipore). Solutions were thoroughly degassed by bubbling nitrogen (oxygen free) for at least 30 mins. Electrochemical experiments were carried out in a thermostated Faraday cage at 298 ($\pm$ 1) K, using a PalmSens Bipotentiostat, (Palm Instruments BV, Netherlands). The platinum interdigitated ring/band working electrodes were fabricated by Micrux Technologies (Spain) using photolithographic techniques on a pyrex substrate. Platinum reference and counter electrodes are integrated by photolithography onto this substrate.

\section{Experimental Results}

\subsection{Interdigitated band array}

\begin{figure}[h]
\begin{center}
\includegraphics[width = 0.9\textwidth]{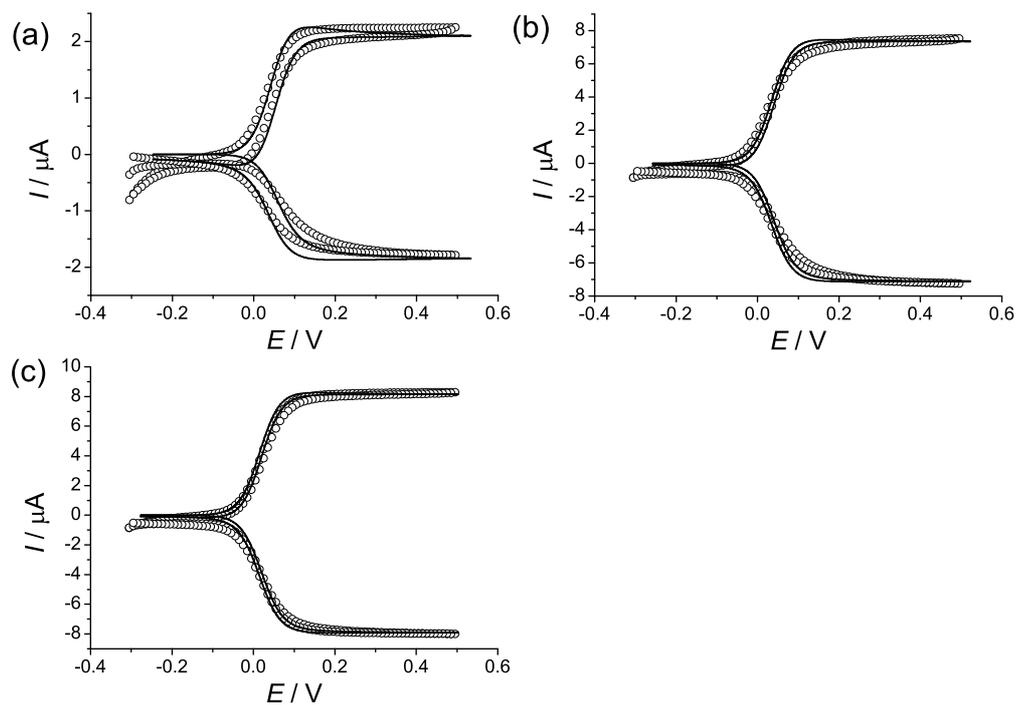}
\caption{Experimental (circles) and simulated (lines) cyclic voltammograms of 1 mM potassium ferrocyanide at IDBA 1 (a), IDBA 2 (b) and IDBA 3 (c) electrodes (see Table \ref{GEOMETRIES}). In all cases the scan rate was 50 mV s$^{-1}$.} \label{IDBA FITS}
\end{center}
\end{figure}

Cyclic voltammetry was carried out on aqueous 1 mM potassium ferrocyanide, K$_4$Fe(CN)$_6$, at a scan rate of 50~mV~s$^{-1}$ in the presence of 0.1 M KCl at three different Pt IDBA electrodes of varying geometries (summarised in Table \ref{GEOMETRIES}). In all three cases the lengths of the electrodes is 2 mm. The results for IDBA1, IDBA2 and IDBA3 are shown in Fig. \ref{IDBA FITS}, along with theoretical best fits. Literature values for the diffusion coefficients\cite{Klymenko2004, Adams1969} of Fe(CN)$_6^{4-}$ and Fe(CN)$_6^{3-}$ of $6.6\times10^{-12}$~m$^2$~s$^{-1}$ and $7.5\times10^{-12}$~m$^2$~s$^{-1}$ respectively were used in the simulations, with a best fit concentration determined to be $0.95 \pm 0.05$ mM. $\alpha$ and $\beta$ are set at 0.5, and $k^0$ is arbitrarily set at 1000~m~s$^{-1}$ to ensure complete electrochemical reversibility. Excellent fits are seen, validating the IDBA model developed above. Collection efficiencies for IDBA1, 2 and 3 at the vertex potential were measured as 79\%, 96\% and 96\%. This is consistent with smaller inter electrode gaps (relative to the electrode widths) producing higher collection efficiencies. 

\subsection{Interdigitated ring array}

\begin{figure}[h]
\begin{center}
\includegraphics[width = 0.9\textwidth]{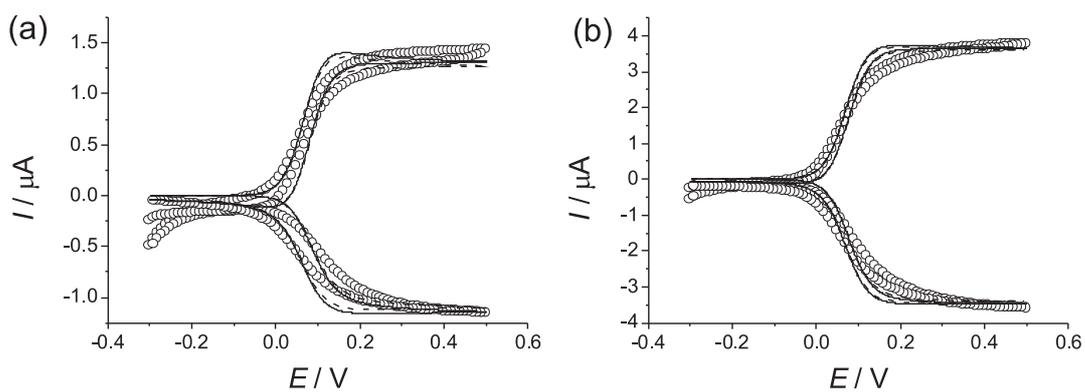}
\caption{Cyclic voltammograms of 1 mM potassium ferrocyanide at IDRA 1 (a) and IDRA 2 (b) electrodes (see Table \ref{GEOMETRIES}). In all cases the scan rate was 50 mV s$^{-1}$. Circles: Experimental data, solid lines: Simulations using the full IDRA model, dashed lines: Simulations approximating the rings as bands.} \label{IDRA FITS}
\end{center}
\end{figure}

The cyclic voltammetry of aqueous 1 mM K$_4$Fe(CN)$_6$ was then repeated on two Pt IDRA electrodes (geometries summarised in Table \ref{GEOMETRIES}), again at a scan rate of 50~mV~s$^{-1}$, in the presence of 0.1 M KCl. The experiments were simulated using the full IDRA model outlined above, and using the IDBA model with each pair of generator-collector electrodes having an area equal to each of the generator electrodes in the IDRAs. The results are shown in Fig. \ref{IDRA FITS}. Collection efficiencies at the potential vertex were measured as 78\% and 92\% for IDRA1 and 2 respectively, again relfective of the smaller gap size in IDRA2. For the simulations, the same literature diffusion coefficients as above were used, and in both cases the concentration of K$_4$Fe(CN)$_6$ was fixed at 1 mM. It is seen that the two models produce very similar results, which both fit well with the experimental data, with differences likely being due to capacitative effects. The experimentally measured currents are dominated by the larger, more peripheral rings, which have small enough curvature for radial diffusion to be negligible on these timescales. For IDRA1, the outermost electrode pair has an $R_0$ value of 46. The equation for $R_{5\%}$ developed above gives an $R_{5\%}$ value of 42.16 for $d^{'} = 1.0$, meaning this outer electrode pair can be reasonably well modelled as bands. For IDRA2, the outer electrdoe pair has $R_0 = 94.3$, well above the $R_{5\%}$ value of 33.18 for this geometry. This allows us to model these IDRAs using the computationally more simplistic IDBA model.

\section{Conclusions}

In this study, interdigitated band and ring electrodes were simulated and theoretical results compared. It was found that if a generator-collector pair of electrodes in a ring array has a large enough radius, the pair can be approximated as a pair of parallel linear bands. An equation has been developed for what radius this pair in a ring array needs to be to produce a result within 5\% of that at parallel bands: $R_0 \geq 22.4d^{'} + 19.7$ in the range $0.1 \leq d^{'} \leq 1.0$. This was found to be independant of the scan rate used. The models for interdigitated bands and rings were used to simulate experimental data for the oxidation of Fe(CN)$_6^{4-}$, and found to give good fits. It was also shown that if the inequality given above is satisfied for the outermost pair of electrodes in a ring array, then the array can be effectively modelled using the more simplistic interdigitated band array model.

\section*{Acknowledgments}

We thank EPSRC for financial support. EOB and GEML further than St John's College, Oxford and NERC for additional support.

\clearpage

\providecommand*\mcitethebibliography{\thebibliography}
\csname @ifundefined\endcsname{endmcitethebibliography}
  {\let\endmcitethebibliography\endthebibliography}{}

\clearpage

\section*{Figures}

\clearpage

\clearpage

\clearpage

\clearpage

\clearpage

\clearpage

\clearpage

\clearpage

\clearpage

\section*{Tables}

\clearpage

\begin{table}
\begin{center}
\begin{tabular}{l l l}
\hline
Parameter & Description & Units \\
\hline
$\alpha$ & Transfer coefficient & Unitless\\
\\
$c_\mathrm{i}$ & Concentration of species i & mol m$^{-3}$\\
\\
$c_\mathrm{i}^{*}$ & Bulk solution concentration of species i & mol m$^{-3}$\\
\\
$c_\mathrm{i}^{0}$ & Concentration of species i at electrode surface & mol m$^{-3}$\\
\\
$D_\mathrm{i}$ & Diffusion coefficient of species i & m$^2$ s$^{-1}$ \\
\\
$E$ & Applied potential & V\\
\\
$E_f$ & Formal potential of A/B couple & V\\
\\
$F$ & Faraday constant = 96485 & C mol$^{-1}$ \\
\\
$I$ & Current & A\\
\\
$k^0$ & Electrochemical rate constant & m s$^{-1}$\\ 
\\
$l$ & Electrode length & m \\
\\
$\nu$ & Scan rate & V s$^{-2}$\\
\\
$R$ & Gas constant = 8.314 & J K$^{-1}$ mol$^{-1}$\\ 
\\
$r$ & Radial coordinate in cylindrical space & m \\
\\
$r_0$ & Radius to centre of generator electrode in cylindrical & \\
& simulation space & m \\
\\
$T$ & Temperature & K\\
\\
$t$ & time & s\\
\\
$w_e$ & width of electrode & m \\
\\
$w_g$ & width of inter electrode gap & m \\
\\
$x$ & $x$ coordinate in Cartesian space& m \\
\\
$y$ & $y$ coordinate in Cartesian space& m \\
\\
$z$ & $z$ coordinate in cylindrical space& m \\ 
\hline
\end{tabular}
\end{center}
\caption{List of symbols}
\label{DIMENSIONAL}
\end{table}

\clearpage

\begin{table}
\begin{center}
\begin{tabular}{c c}
Dimensionless Parameter & Definition \\
\hline
\\
$C_\mathrm{i}$ & $\frac{c_\mathrm{i}}{c_\mathrm{A}^*}$ \\
\\
$d^{'}$ & $\frac{w_g}{w_e}$ \\
\\
$D_\mathrm{i}^{'}$ & $\frac{D_\mathrm{i}}{D_\mathrm{A}}$ \\
\\
$j$ (bands) & $\frac{I}{nFlDc_\mathrm{A}^*}$ \\
\\
$j$ (rings) & $\frac{I}{2\pi nFr_0Dc_\mathrm{A}^*}$ \\
\\
$K^0$ & $\frac{w_e}{D_\mathrm{A}}k^0$ \\
\\
$Q$ & $\frac{I_\mathrm{ring}^{\mathrm{peak}}}{I_\mathrm{band}^{\mathrm{peak}}}$ \\
\\
$R$ & $\frac{r}{w_e}$ \\
\\
$\sigma$ & $\frac{Fw_e^2}{RTD_\mathrm{A}}\nu$ \\
\\
$\theta$ & $\frac{RT}{F}E$ \\
\\
$\theta_f^\minuso$ & $\frac{RT}{F}E_f^\minuso$ \\
\\
$\tau$ & $\frac{D_\mathrm{A}}{w^\mathrm{2}_\mathrm{e}}t$ \\
\\
$X$ & $\frac{x}{w_e}$ \\
\\
$Y$ & $\frac{y}{w_e}$ \\
\\
$Z$ & $\frac{z}{w_e}$ \\
\hline
\end{tabular}
\end{center}
\caption{Normalised parameters. Species A refers to the species initially present in solution before the experiment/simulation begins.}
\label{DIMENSIONLESS}
\end{table}

\clearpage

\begin{table}
\begin{center}
\begin{tabular}{l l l l}
\hline
$\tau$ & $X$ & $Y$ & Boundary condition(s)\\
\hline
$\tau < 0$ & All $X$ & All $Y$ & $C_\mathrm{A} = 1$, $C_\mathrm{B} = 0$\\
\\
$\tau \geq 0$ & $-0.5 - \frac{1}{2}d < X \leq - \frac{1}{2}d$ & $Y = 0$ & $D^{'}_\mathrm{A}\left(\frac{\partial{C_\mathrm{A}}}{\partial{Y}}\right) = K^0C_\mathrm{A}^0\text{exp}\left[-\alpha\left(\theta - \theta_f^\minuso\right)\right]$\\
& & &$\quad\quad\quad\quad\quad\quad - K^0C_\mathrm{B}^0\text{exp}\left[\left(1-\alpha\right)\left(\theta - \theta_f^\minuso\right)\right]$\\
\\
& & & $D^{'}_\mathrm{B}\left(\frac{\partial{C_\mathrm{B}}}{\partial{Y}}\right) = -D^{'}_\mathrm{A}\left(\frac{\partial{C_\mathrm{A}}}{\partial{Y}}\right)$\\
\\
& $\frac{1}{2}d \leq X < \frac{1}{2}d + 0.5$ & $Y = 0$ & $D^{'}_\mathrm{A}\left(\frac{\partial{C_\mathrm{A}}}{\partial{Y}}\right) = -D^{'}_\mathrm{B}\left(\frac{\partial{C_\mathrm{B}}}{\partial{Y}}\right)$\\
\\
& & & $C_\mathrm{B} = 0$\\
\\
& $X = -0.5 - \frac{1}{2}d$ & All $Y$ & $\frac{\partial{C_\mathrm{i}}}{\partial{X}} = 0$\\
\\
& $X = 0.5 + \frac{1}{2}d$ & All $Y$ & $\frac{\partial{C_\mathrm{i}}}{\partial{X}} = 0$\\
\\
& All $X$ & $Y = Y_\mathrm{max}$ & $\frac{\partial{C_\mathrm{i}}}{\partial{Y}} = 0$\\
\hline
\end{tabular}
\end{center}
\caption{Normalised boundary conditions for the IDBA model}
\label{IDA BOUNDARY CONDITIONS}
\end{table}

\clearpage

\begin{table}
\begin{center}
\begin{tabular}{l l l l}
\hline
$\tau$ & $R$ & $Z$ & Boundary condition(s)\\
\hline
$\tau < 0$ & All $R$ & All $Z$ & $C_\mathrm{A} = 1$, $C_\mathrm{B} = 0$\\
\\
$\tau \geq 0$ & $R_0 - 0.5 < R \leq R_0 + 0.5$ & $Z = 0$ & $D^{'}_\mathrm{A}\left(\frac{\partial{C_\mathrm{A}}}{\partial{Z}}\right) = K^0C_\mathrm{A}^0\text{exp}\left(-\alpha\left(\theta - \theta_f^\minuso\right)\right)$\\
& & &$\quad\quad\quad\quad\quad\quad - K^0C_\mathrm{B}^0\text{exp}\left[\left(1-\alpha\right)\left(\theta - \theta_f^\minuso\right)\right]$\\
\\
& & & $D^{'}_\mathrm{B}\left(\frac{\partial{C_\mathrm{B}}}{\partial{Z}}\right) = -D^{'}_\mathrm{A}\left(\frac{\partial{C_\mathrm{A}}}{\partial{Z}}\right)$\\
\\
& $R_0 - 0.5 - d \leq R \geq R_0 + 0.5 + d$ & $Z = 0$ & $D^{'}_\mathrm{A}\left(\frac{\partial{C_\mathrm{A}}}{\partial{Z}}\right) = -D^{'}_\mathrm{B}\left(\frac{\partial{C_\mathrm{B}}}{\partial{Z}}\right)$\\
\\
& & & $C_\mathrm{B} = 0$\\
\\
& $X = R_0 - 1 - d$ & All $Z$ & $\frac{\partial{C_\mathrm{i}}}{\partial{X}} = 0$\\
\\
& $X = 0.5 + 1 + d$ & All $Z$ & $\frac{\partial{C_\mathrm{i}}}{\partial{X}} = 0$\\
\\
& All $R$ & $Z = Z_\mathrm{max}$ & $\frac{\partial{C_\mathrm{i}}}{\partial{Z}} = 0$\\
\hline
\end{tabular}
\end{center}
\caption{Normalised boundary conditions for the IDRA model}
\label{IDRA BOUNDARY CONDITIONS}
\end{table}

\clearpage

\begin{table}
\begin{center}
\begin{tabular}{l l l l l}
\hline
Electrode & Bands/rings & Electrode width / $\mu$m & Gap width / $\mu$m & Number of pairs\\
\hline
IDBA 1 & Bands & 10 & 10 & 15 \\
IDBA 2 & Bands & 5 & 2 & 42 \\
IDBA 3 & Bands & 3 & 2 & 59 \\
IDRA 1 & Rings & 10 & 10 & 12 \\
IDRA 2 & Rings & 5 & 3 & 30 \\
\hline
\end{tabular}
\end{center}
\caption{Summary of geometries of electrode arrays used in this study. All the band electrodes are 2 mm in length.}
\label{GEOMETRIES}
\end{table}

\end{document}